\documentclass[proceedings]{rmaa}
\newcommand\I{{\'\i}}
\newcommand\HII{H\,{\sc ii} }

\title{PN-Carbon Yields and the Chemical Evolution of the Galaxy}
\author{Leticia Carigi 
  \affil{Instituto de Astronom\'{\i}a, Universidad Nacional Aut\'onoma de M\'exico } }

\fulladdresses{
\item Leticia Carigi: Instituto de Astronom\'{\i}a,
  UNAM, Apartado Postal 70-264, 04510 M\'exico D.F.,
   M\'exico

(carigi@astroscu.unam.mx) }

\shortauthor{Carigi}
\shorttitle{PN-Carbon Yields}


\abstract{
Two sets of observational carbon stellar yields for low-and-intermediate mass stars 
are computed from
 planetary nebulae abundances  derived from C II $\lambda4267$ and 
C III $\lambda\lambda1906+1909$ lines, respectively.
By comparing C/O values observed in stars of the solar vicinity
and Galactic \HII regions with those predicted by chemical evolution models 
for the Galaxy, which assume these observational yields,
I conclude that the C abundances derived from permitted lines are better
than those derived from forbidden lines. 
  }

\resumen{
Calculo dos conjuntos de rendimientos estelares observacionales
de carbono 
 para estrellas de masa  intermedia y baja
basados en  abundancias derivadas a partir de l\I neas permitidas
y prohibidas de C II y C III ($\lambda4267$ y $\lambda\lambda1906+1909$, respectivamente).
Comparando los valores de C/O observados en la vecindad solar y en regiones \HII
Gal\'acticas con aquellos predichos por modelos de evolucion qu\I mica para la 
Galaxia que suponen estos rendimientos observacionales, 
se concluye que las abundancias de C basadas en 
l\I neas permitidas son mejores que aquellas derivadas a partir de l\I neas
prohibidas. 
 }
\indexauthor{Carigi, L.}

\begin{document}

\maketitle

\section{Introduction}
\label{sec:intro}

The $N({\rm C}^{++})/N({\rm H}^+)$ values derived from the permitted line (PL)
C II $\lambda4267$ are higher, by as much as a factor of 10, than those determined from the forbidden lines (FL)
C III $\lambda \lambda1906+1909$.
Several explanations for this discrepancy have been presented in the literature,
but the problem remains open.
Since PNe are important for the C enrichment of the interstellar medium, 
a successfull chemical evolution model for the solar vicinity and the Galactic
disk (Carigi 2000)
is used 
to discriminate between the PN-C abundances derived from permitted lines
and the PN-C abundances obtained from forbidden lines.

\section{PN Observational Yields} 
\label{sec:constraints}

 Average (C/H)$_{\rm PN}^{\rm PL}$ and (C/H)$_{\rm PN}^{\rm FL}$ values are computed from
15 type I PNe (PNI), and from 21 type II and III PNe (PNII/III).
$<$(C/H)$_{\rm PNI}^{\rm PL}>$ is calculated from the $N({\rm C}^{++})/N({\rm O}^{++}$) 
and $N({\rm O})/N({\rm H})$ values
given by Peimbert et al. (1995a). 
The $<$(C/H)$_{\rm PNI}^{\rm FL}>$ value 
is obtained from $<$(C/H)$_{\rm PNI}^{\rm PL}>$ and the average 
of $N({\rm C}^{++})/N({\rm H}^+)_{\rm PL}/N({\rm C}^{++})/N({\rm H}^+)_{\rm FL}$ 
ratios taken from Peimbert et al. (1995b).
The $<$(C/H)$_{\rm PNII/III}^{\rm PL}>$  and  $<$(C/H)$_{\rm PNII/III}^{\rm FL}>$ values are computed from 
the $N({\rm C}^{++})/N({\rm H}^+)$ given by Peimbert et al. (1995b) and corrected for the contribution of
$N({\rm C}^+)/N({\rm H}^+)$ by adding 0.1 dex.

According to stellar evolution models, the PNI progenitors are stars with initial mass between
2.4 and 8 $M_\odot$ and the PNII/III  progenitors are stars with  $0.8<m/M_\odot<2.4$.
The C yields for PN progenitors (C$_{\rm PN}$ yields)
are calculated based on the $<$(C/H)$_{\rm PN}>$ from permitted lines and forbidden lines,
neglecting the ejected mass by winds and assuming that  the $<$(C/H)$_{\rm PN}>$ values 
are independent of the initial metallicity of the progenitors. 
I assume that the O yields for PN progenitors are null.
The mass ejected by PN progenitors are taken from van den Hoek \& Groenewegen (1997).

\section{Models }
\label{sec:models}

 All models are built to reproduce 
the observed gas fraction distribution of the Galaxy
and the observed O/H  Galactic gradient from 4 to 10 kpc.
Models adjust
the rise of C/O with metallicity
in the solar neighborhood shown by dwarf stars, the Sun, and Orion; and
the decrease of the C/O abundance with Galactocentric
 distance
derived from the Galactic \HII regions M17, M8,
 and Orion.

The models are very similar to the infall model of Carigi (2000),
but 
in this work there are some differences in the assumptions about stellar yields:

a) Just one set of metal-dependent stellar yields from massive stars
($8<m/M_\odot<120$) is considered:
Geneva yields (Maeder 1992).
Carigi  concluded that
only models with the Geneva yields can 
reproduce both,
the increase of C/O with $Z$ in the solar vicinity
and the negative C/O gradient. 

b) Four sets of stellar yields for low and intermediate mass stars
($0.8<m/M_\odot<8$)  are used:
i) two  metal-dependent-theoretical ones:
van den Hoek \& Groenewegen (1997, Amsterdam yields),  
and Marigo et al. (1996, 1998)  and Portinari et al. (1998) (Padova yields);
ii) two metal-independent-observational stellar yields:
PN yields from permitted  lines and  PN yields from forbidden lines.

C/O depends on
the initial mass function and
the C and O yields.
Since the initial mass function and the massive-star yields are fixed,
the C/O value is used to test the different sets of C and O yields for 
low and intermediate mass stars

\section{Results and Conclusions}
\label{sec:conclusion}

By comparing  observed and theoretical C yields, it can be noted that:
i) there is a very good agreement between the ${\rm C^{PL}_{PN}}$ yields  and those
predicted
by stellar evolution models with $Z=Z_\odot$, and
ii) the ${\rm C^{FL}_{PN}}$ yields  are lower, by as much as a factor of 4, than
those predicted
by stellar evolution models with $Z=Z_\odot$.

\medskip
From chemical evolution models of the Galaxy, 
I conclude that:

a) Models with C$_{\rm PN}$ yields derived from permitted lines
 match
all the observational constraints,
in particular  they reproduce
the C/O absolute values observed in stars of the solar vicinity
and in \HII regions of the Galactic disk.

b) Models with C$_{\rm PN}$ yields from forbidden lines
fail to  reproduce the
C/O ratios in dwarf stars of different ages  in the solar vicinity,
the Sun,
and M17.

c) Models with C$_{\rm PN}$ yields derived from permitted lines
agree
with models based on theoretical yields,
in particular
showing better agreement
with models based on the Padova yields  than
models based on the Amsterdam yields.

d) The C/O values predicted with ${\rm C_{PN}}$ yields
derived from permitted lines are 0.1 dex higher
 than those
obtained with the Amsterdam yields, and 0.05 dex lower
than those computed with the Padova yields.

\end{document}